\documentclass[a4paper,twocolumn,11pt]{quantumarticle}
\pdfoutput=1
\usepackage[utf8]{inputenc}
\usepackage[english]{babel}
\usepackage[T1]{fontenc}
\usepackage{amsmath}
\usepackage{hyperref}
\usepackage{fancyhdr,graphicx,amsmath,amssymb}
\usepackage[ruled,vlined]{algorithm2e}
\usepackage{graphicx}% Include figure files
\usepackage{subcaption}
\usepackage[numbers]{natbib}
\usepackage{tikz}
\usepackage{lipsum}

\begin{document}

\title{Preparing random state for quantum financing with quantum walks}

\author{Yen-Jui Chang}

\affiliation{Department of Physics, National Taiwan University, Taipei 106216, Taiwan}
\affiliation{NTU-IBM Quantum Hub, National Taiwan University, Taipei 106216, Taiwan} 
\orcid{0009-0009-7400-9183}
\author{Wei-Ting Wang}
\affiliation{Department of Physics, National Taiwan University, Taipei 106216, Taiwan
}
\author{Hao-Yuan Chen}
\affiliation{Department of Computer Science and Information Engineering, Tamkang University, New Taipei City 251301, Taiwan
}
\author{Shih-Wei Liao}
\affiliation{Department of Computer Science and Information Engineering, National Taiwan University, Taipei 106216, Taiwan
}
\author{Ching-Ray Chang}
\email{crchang@phys.ntu.edu.tw}
\affiliation{Department of Physics, National Taiwan University, Taipei 106216, Taiwan}
\affiliation{Quantum Information Center, Chung Yuan Christian University, Taoyuan City 320314, Taiwan}

\thanks{You can use the \texttt{email} \texttt{thanks} commands to add additional information for the preceding \texttt{author}.}

\maketitle

\begin{abstract}
In recent years, there has been an emerging trend of combining two innovations in computer science and physics to achieve better computation capability. Exploring the potential of quantum computation to achieve highly efficient performance in various tasks is a vital development in engineering and a valuable question in sciences, as it has a significant potential to provide exponential speedups for technologically complex problems that are specifically advantageous to quantum computers. However, one key issue in unleashing this potential is constructing an efficient approach to load classical data into quantum states that can be executed by quantum computers or quantum simulators on classical hardware. Therefore, the split-step quantum walks (SSQW) algorithm was proposed to address this limitation. We facilitate SSQW to design parameterized quantum circuits (PQC) that can generate probability distributions and optimize the parameters to achieve the desired distribution using a variational solver. A practical example of implementing SSQW using Qiskit has been released as open-source software. Showing its potential as a promising method for generating desired probability amplitude distributions highlights the potential application of SSQW in option pricing through quantum simulation. 
\end{abstract}

\section{\label{sec:level1}Introduction}

In quantum mechanics, particles have a wave-like nature, meaning their position and momentum cannot be known simultaneously with arbitrary precision. This uncertainty is a result of the wave-like nature of quantum particles and is known as Heisenberg's Uncertainty Principle\cite{Heisenberg1930}. This wave-like nature of quantum particles also allows them to exist in a superposition of states, where a particle is in multiple states simultaneously. The wave-like nature of quantum particles also leads to the phenomenon of quantum interference, where the probability of a particle being in a certain state is determined by the sum of the probabilities of all the possible paths it could take to reach that state. Additionally, the wave-like nature of quantum particles also gives rise to quantum entanglement, where the properties of two or more quantum particles are correlated in such a way that the state of one particle cannot be described independently of the state of the other particle. This can lead to correlations that cannot be explained by classical physics and are often considered random or unpredictable. The wave-like nature of quantum particles is closely related to the randomness in quantum dynamics; it gives rise to the uncertainty principle, wave-particle duality, quantum interference, superposition of states, and entanglement that lead to the unpredictability and randomness of quantum systems. 

Recently, the randomness of quantum systems has been demonstrated \cite{Choi2023} the distribution of states close to the Haar-random. Haar-random\cite{Nielsen2000,Guise2018} is particularly important in the field of quantum computing, as they can be used to generate arbitrary quantum states, perform quantum gates and perform quantum walk that enables applications of this quantum computations in a much wider context.
Quantum computers utilize quantum-mechanical phenomena, such as superposition and entanglement, to perform operations on data. Unlike classical computers, which encode data into bits that are either 0 or 1. Due to the unique properties of quantum mechanics, quantum bits(qubits) exist in multiple states simultaneously and are strongly correlated with one another. This makes it possible for a quantum computer to perform certain operations, such as factorization\cite{Shor1994} and simulation\cite{Feynman1982,Trabesinger2012}, much more quickly than a classical computer. The achievement of quantum advantage is considered an important goal in the field of quantum computing. The three steps of realizing quantum advantage: loading quantum state, quantum algorithm computation, and result measurements, are shown in Fig.1. 

\begin{figure}[htb]
\centering
\includegraphics[width=8.5cm]{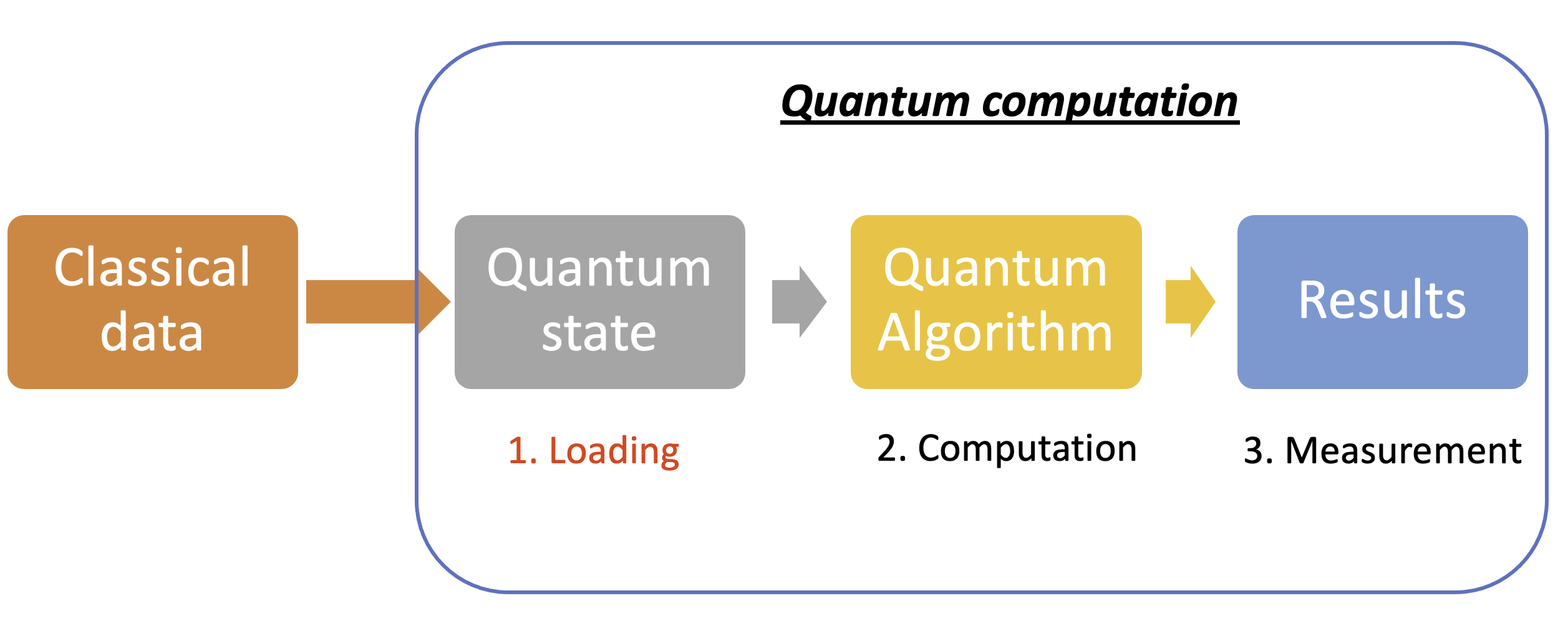}
\caption{ The three generic steps to realize the quantum advantage. }
\label{fig:FIG1}
\end{figure}

This allows quantum computers to have the potential to solve some challenging problems considerably faster than classical computers using the most efficient known quantum algorithms\cite{HHL:2009,PCA:2014,Brassard_2002,QSVM:2019,QC2021,Montanaro2016}, provided that the data can be efficiently loaded into a quantum state. Recently, there has been an increased focus on utilizing quantum computers to generate specific probability distributions with the quantum nature of randomness. The ability to generate probability distributions in a quantum circuit has many conceivable applications. One area where this has been used in finance\cite{9222275,Herman2022,Rebentrost2018} is assigning probability amplitudes to a quantum system's basis states. With n qubits, a system would have $2^{n}$ basis states. The generator function is a probability amplitude generator, which can be modified to achieve the desired probability amplitude distribution. The generator must designate amplitudes to these states in a way that closely resembles a targeted probability distribution that can create a quantum superposition state. 
\begin{equation} \label{eq:superposition}
\;|\psi\rangle= \sum_{i} \sqrt{{p_{i}}} \;|i\rangle 
\end{equation}
where $\;|i \rangle$ are orthonormal quantum states. Efficient methods for generating probability distributions in quantum circuits are an important study area. Additionally, generating probability distributions is also useful in creating initial states for Hamiltonian simulation, simulating a wave function's time evolution. 

 An ineffective and slow data loading method could diminish the potential impact of quantum computing\cite{PhysRevLett.122.020502}. The preparation methods of a generic quantum state have been developed\cite{Grover:loading2002,Zoufal:qGAN2019,Kalayn:Loading2022}.   Quantum Generative Adversarial Networks (qGAN) have been demonstrated to load distributions\cite{Zoufal:qGAN2019}. The qGAN utilizes a combination of a quantum generator and a classical discriminator to learn the probability distribution of classical training data. The quantum generator, a parametrized quantum channel, is trained to convert an input state of n-qubits, represented by $\;|\psi\rangle$, into an output state of n-qubits. The earliest papers on generating probability distributions had been proposed\cite{Grover:loading2002}. A scheme shows how to generate a superposition of quantum states by taking an ancilla register that performs a controlled rotation of angle $\theta_{i}$:
\begin{equation} \label{eq:controlratation}
\sqrt{{p_{i}}} \;|i\rangle \to \sqrt{{p_{i}}} \;|i\rangle \otimes (\cos \theta\;|0\rangle + \sin \theta |1\rangle)
\end{equation}
Recently, a method using variational solvers to fix the rotation parameters of the gate has been proposed to generate symmetrical and asymmetrical probability distributions\cite{Kalayn:Loading2022}. The authors demonstrated trajectories of the individual quantum states to understand the effect of an ancilla register to control rotation. 

In this paper, we present a method, based on eq.(\ref{eq:controlratation}), for incorporating SSQW into probability distributions. The structure of the paper is as follows. First, we begin by reviewing the theoretical foundations of quantum walks and explain their role that quantum walks for loading probability distributions. Then, we introduce the methodology of SSQW and present the SSQW-based loading method on different test cases, including normal, log-normal, and real stock price distributions, with a quantum simulator accessible via the IBM Q Experience. We then demonstrate using the SSQW-based loading method to facilitate quantum benefit in financial derivative pricing. Finally, we provide conclusions and discuss open questions and potential additional applications of the scheme.

\section{Quantum Walk}

Quantum walks (QW)\cite{Feynman:quantumcomputers,Aharonov1993,Childs:quantumwalk,Mallick:DCA2016,Rajauria:DTQW,Wanzambi:CoinedQuantumWalks,Venegas_Andraca_2012,10.1145/780542.780552}, which are the quantum equivalent of classical random walks, are used as a foundation for generating models of controlled quantum simulation. Quantum walks enable the walker to simulate several quantum mechanical phenomena by tuning the parameters and evolution coin operators of a QW. The formalism for quantum walks is broadly classified into the discrete-time quantum walk (DTQW) and the continuous-time quantum walk (CTQW). Both approaches have unique features that cause them suitable for performing quantum computing tasks. Here we will focus only on the one-dimensional DTQW. A classical walk can be described using just a position Hilbert space, while a DTQW, or quantum walk, requires an additional coin Hilbert space to express its dynamics fully. This coin space represents the internal state of the walker and is necessary to capture the controlled dynamics of the walker.
Hilbert space of quantum walk is defined as follows. 

\begin{equation} \label{eq:HilbertSpace}
\mathcal{H} = \mathcal{H}_{c}  \otimes \mathcal{H}_{p}
\end{equation}
where $\mathcal{H}_{c}$ is the coin Hilbert space and  $\mathcal{H}_{p}$ is the position Hilbert space. The coin Hilbert space for one-dimensional DTQW has the basis states $\{\;|\uparrow\rangle = \begin{pmatrix}
  1\\ 
  0
\end{pmatrix} , \;|\downarrow\rangle = \begin{pmatrix}
  0\\ 
  1
\end{pmatrix}\}$ and the position Hilbert space is defined by the basis states $\;|x\rangle $ where $x\in Z$. The probability amplitude of the quantum state at position $x$ can be represented by

\begin{equation} \label{eq:amplitudeofthequantumstate}
\Psi(x,t) \rangle =
\begin{pmatrix} 
\Psi^{\uparrow} (x,t) \\ \Psi^{\downarrow} (x,t) 
\end{pmatrix}	
\end{equation}
where describe the state of DTQW with two internal degrees of freedom $\{\;|\uparrow\rangle , \;|\downarrow\rangle \}$.
In a Discrete-Time Quantum Walk (DTQW), the system's evolution is governed by two unitary operators: the coin operator and the shift operator. The shift operator moves the walker in a superposition of position states, while the coin operator acts on the coin Hilbert space and determines the amplitudes of the position space. The coin Hilbert space represents the internal state of the walker and plays a crucial role in determining the overall dynamics of the system. A universal operator is defined as
\begin{equation} \label{eq:Coinoperator}
   \hat{C} (\theta, \phi ,\lambda) =
   \begin{pmatrix} 
   \cos(\frac{\theta}{2}) & -e^{i \lambda}\sin(\frac{\theta}{2}) \\ e^{i \phi}\sin(\frac{\theta}{2}) &  e^{i(\lambda+\phi)}\cos(\frac{\theta}{2})
   \end{pmatrix}	
\end{equation}
where are the three independent parameters and the most general unitary coin operator. Therefore, accurately estimating the coin parameters is essential for effectively using quantum walks as a quantum simulation tool and for further research on modeling realistic dynamics. Finding patterns in complex data can be challenging, but an algorithm that automates the learning process can solve this problem.

\begin{figure}[htb]
\centering
\subcaptionbox{}{\includegraphics[width=5cm]{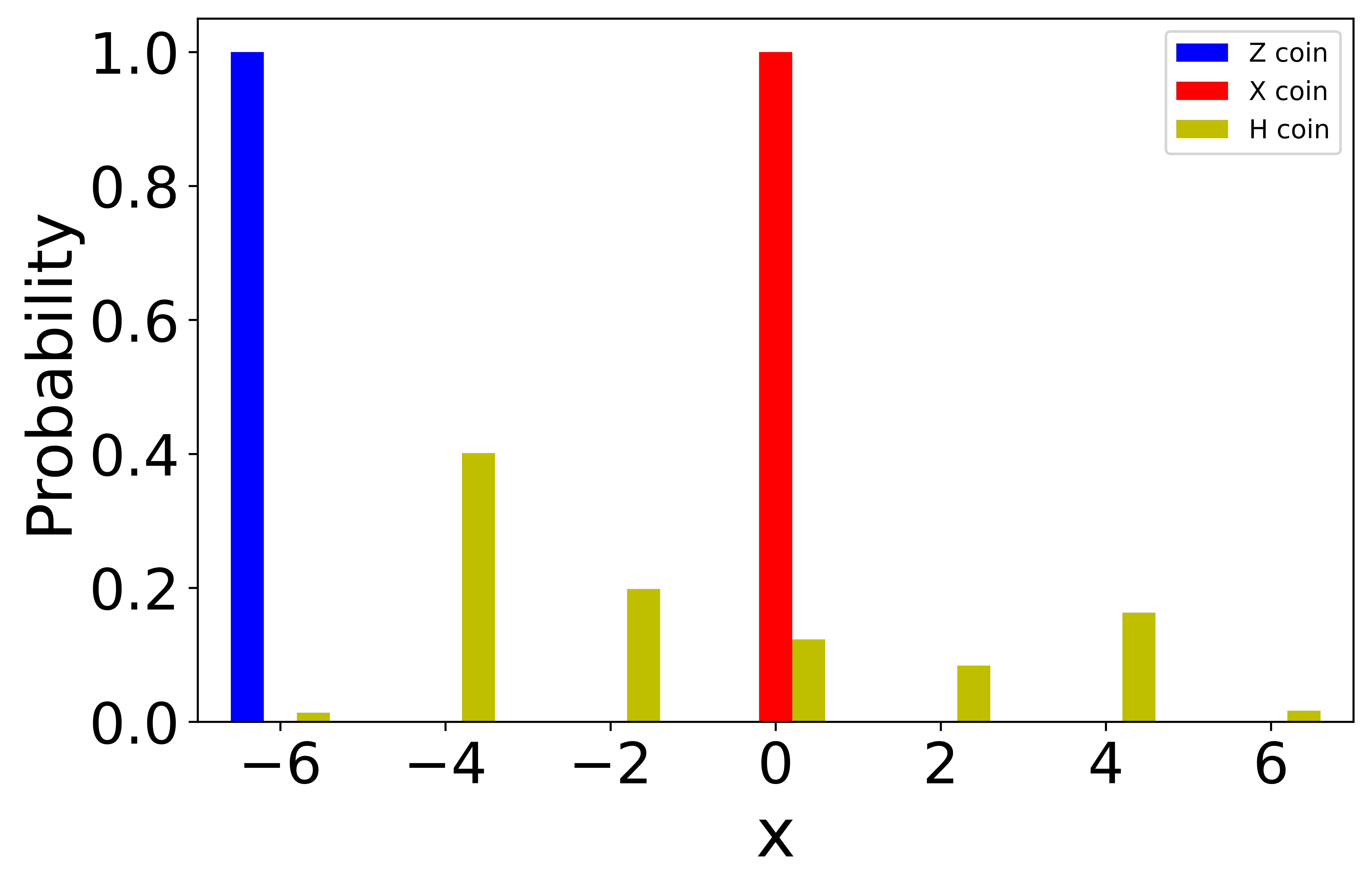}}
\hfill
\subcaptionbox{}{\includegraphics[width=5cm]{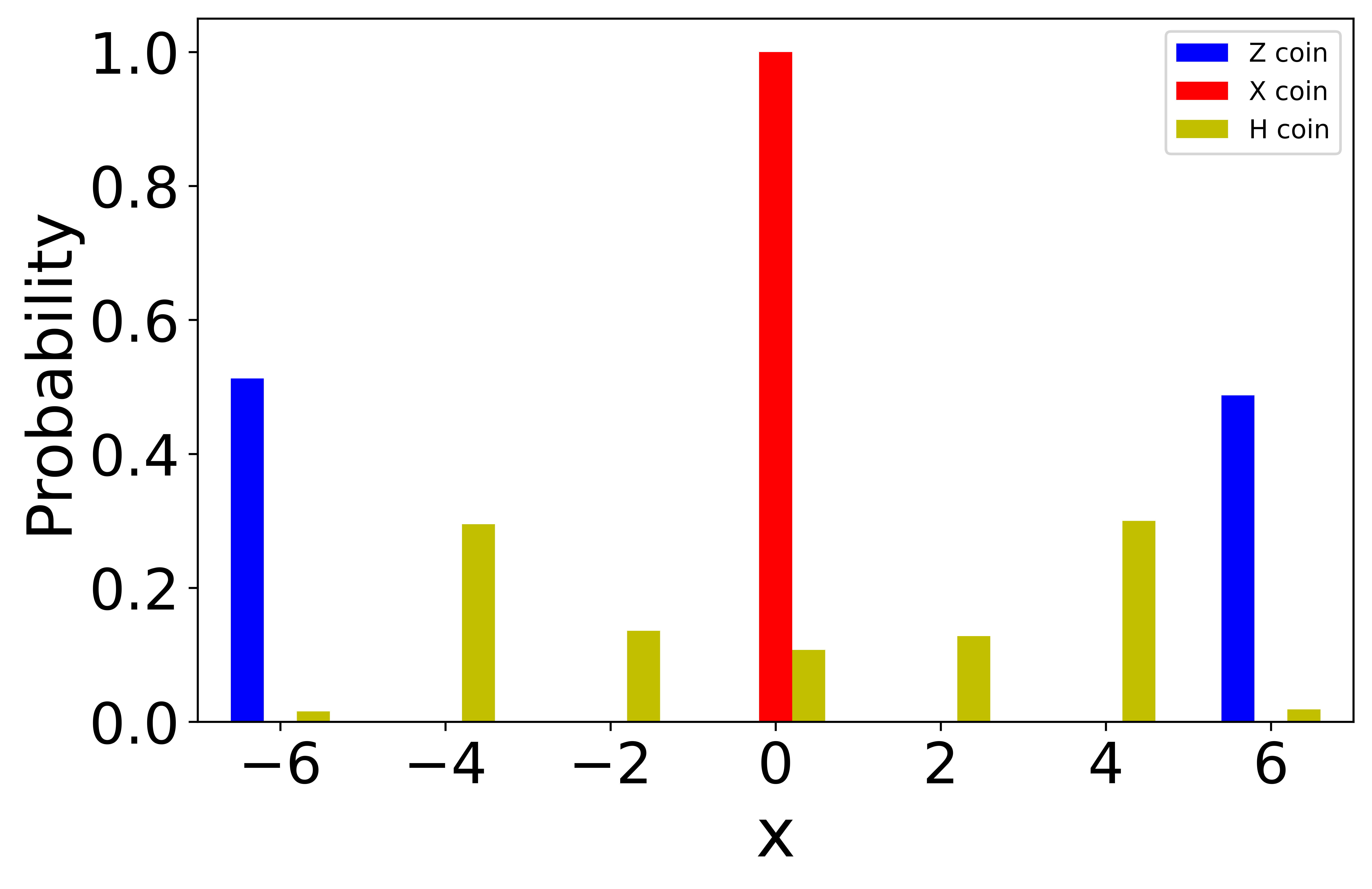}}
\hfill
\subcaptionbox{}{\includegraphics[width=5cm]{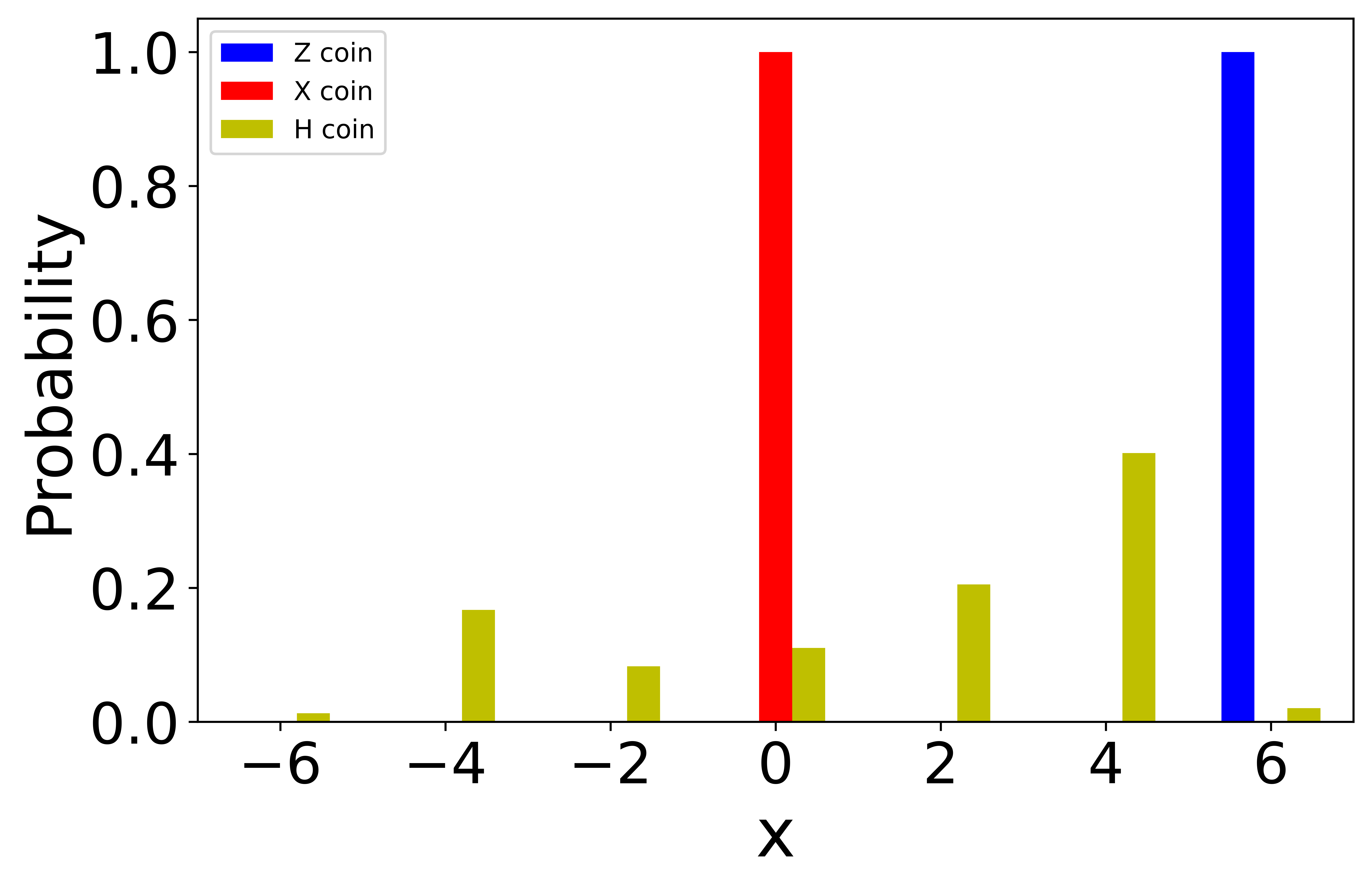}}
\caption{ Demonstrating probability distributions in position space using a Discrete Time Quantum Walk (DTQW) with Z, X, and H gates as coin operators and various initial states (a) Initial state $= |0 \rangle$ (b) Initial state $= \frac{1}{\sqrt{2}}(|0 \rangle+i |1 \rangle )$(c) Initial state $= |1 \rangle$ }
\label{fig:FIG2}
\end{figure}

The shift operator is an essential part of the DTQW. A unitary operator moves the quantum walker in a superposition of position states. The shift operation is defined as,
\begin{equation} \label{eq:shitfoperator}
  \hat{S} = |\downarrow\rangle \langle\downarrow|  \otimes \sum_{x} |x-1\rangle \langle x| + |\uparrow\rangle \langle \uparrow|  \otimes \sum_{x} |x+1\rangle \langle x| 
\end{equation}
In other words, the shift operator shifts the position of the particle one step to the right if the internal state is $|\uparrow\rangle$ or one step to the left $|\downarrow\rangle$.
The initial state of the system is a superposition of the position states, with the internal state of the particle determined by the coin operator and defined as 
\begin{equation} \label{eq:initialstate}
   |\Psi_{0} \rangle = (\alpha|\uparrow \rangle + \beta|\downarrow \rangle) \otimes |x=0 \rangle 	
\end{equation}
At each time step, the shift operator is applied to the position state after the coin operator is applied to the internal state of the particle. Together, these two operators form the evolution operator of the DTQW, which describes the overall dynamics of the system. This process is repeated several times, and the system's final state is a superposition of position states, with the amplitudes determined by the coin operator.
\begin{equation} \label{eq:QWevole}
  \Psi(x,t)=[\hat{S}(\hat{I} \otimes \hat{C})]^{t} |\Psi_{0} \rangle = \hat{W}^{t} |\Psi_{0} \rangle
\end{equation}
where $\hat{I} =  \sum_{x} |x\rangle \langle x| $ is an identity in space.
 The probability distributions in the position space of a walker performing a DTQW, shown in Fig.(\ref{fig:FIG2}), are not similar to a market probability distribution in everyday life. In the marketplace, prices are typically determined by the interaction of buyers and sellers. The price of a good or service in the market is established through the agreement between buyers and sellers. At a particular price point, buyers can decide whether to purchase, while sellers can decide whether to offer their product. The final price is formed as a result of this process. Inspired by the free market economy, we introduce the split-step quantum walk, a specific type of quantum walk that divides the evolution of the quantum system into two separate stages to simulate the probability distribution. 

The probability distributions in position space of a discrete-time quantum walk (DTQW), as shown in  studies\cite{Mallick:DCA2016,Rajauria:DTQW,Venegas_Andraca_2012}, do not resemble the probability distributions in everyday life. In contrast, in a free market economy, the final prices are determined by the agreement between buyers and sellers. Inspired by the dynamics of a free market, the split-step quantum walk (SSQW)\cite{Matsuzawa_2020,Narimatsu:SSQW} is introduced as a way to simulate probability distributions in the quantum realm. SSQW is a specific type of quantum walk that divides the evolution of the quantum system into two steps, one to the right and one to the left. The evolution operator $\hat{W}$ is divided by a composition of two half-steps,

\begin{equation} \label{eq:SSQW_W}
  \hat{W} = \hat{S}_{-}\hat{C}_{\theta_{2}} \hat{S}_{+}\hat{C}_{\theta_{1}} 
\end{equation}
where  $\hat{C}_{\theta_{k}}$ is a universal coin operator as Eq.(\ref{eq:Coinoperator}) and shift operators $\hat{S}_{\pm}$ are defined as,
\begin{equation} \label{eq:SSQW_Shift}
\begin{aligned}
   \hat{S}_{+} &= \sum_{x}[|\uparrow \rangle \langle \uparrow| \otimes |x+1\rangle \langle x| +|\downarrow \rangle \langle \downarrow| \otimes |x\rangle \langle x|  ]  \\
   \hat{S}_{-} &= \sum_{x}[|\uparrow \rangle \langle \uparrow| \otimes |x\rangle \langle x| +|\downarrow \rangle \langle \downarrow| \otimes |x-1\rangle \langle x|  ]
\end{aligned}
\end{equation}
The probability of the walker going right or stopping for a buyer is shown by $\hat{C}_{\theta_{1}}$ while $\hat{C}_{\theta_{2}}$ displays the probability of the walker going left or stopping for a seller. Additionally, $\hat{S}_{+}$ represents the walker goes to right($|\uparrow \rangle$) or stop in place($|\downarrow \rangle$) and $\hat{S}_{-}$ represents the walker goes to left($|\downarrow\rangle$) or stop in place($|\uparrow \rangle$). The concept the SSQW also can be regarded as a method of Haar-random unitary\cite{Choi2023} , which is a collection of unitary operators that can be used to approximate all unitary operators on a given space with high accuracy. 

\begin{figure}[htb]
\centering
\subcaptionbox{}{\includegraphics[width=7cm]{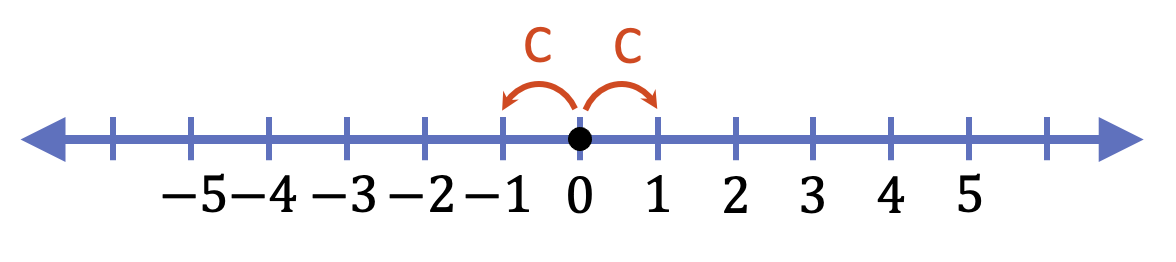}}
\hfill
\subcaptionbox{}{\includegraphics[width=7cm]{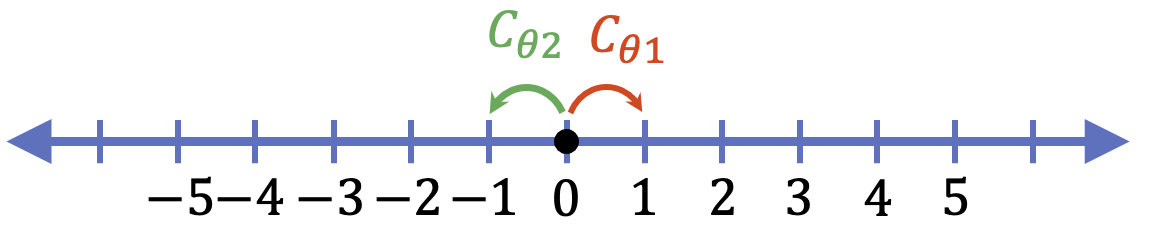}}
\caption{ The scheme of (a)DTQW and (b)SSQW }
\label{fig:FIG3}
\end{figure}

\section{Methods}

In this section, the scheme for the loading probability distribution is introduced with a practical approach to find patterns in complex data and map them to an SSQW dynamic system when loading probability distributions. Also, an SSQW implementation and results evaluation is provided. The architecture of loading the probabilistic data into a quantum state has shown below in Fig.(\ref{fig:FIG4}). Firstly, there are two domains of computation in this scheme of solution. Firstly, a quantum computation domain consists of five qubits for computation, one for coin space and the other for position space. Secondly, a classical optimizer using the Constrained Optimization By Linear Approximation algorithm(COBYLA) is implemented to converge the trained results to the targeted distribution. The classical computation facilitates COBYLA optimizer with a loss function using mean-square error (MSE) to approach the targeted distribution better. Since MSE has a better fitting performance than the L2 loss function, this research primarily set MSE as the loss function of choice.

\begin{figure}[htb]
\centering
\includegraphics[width=7cm]{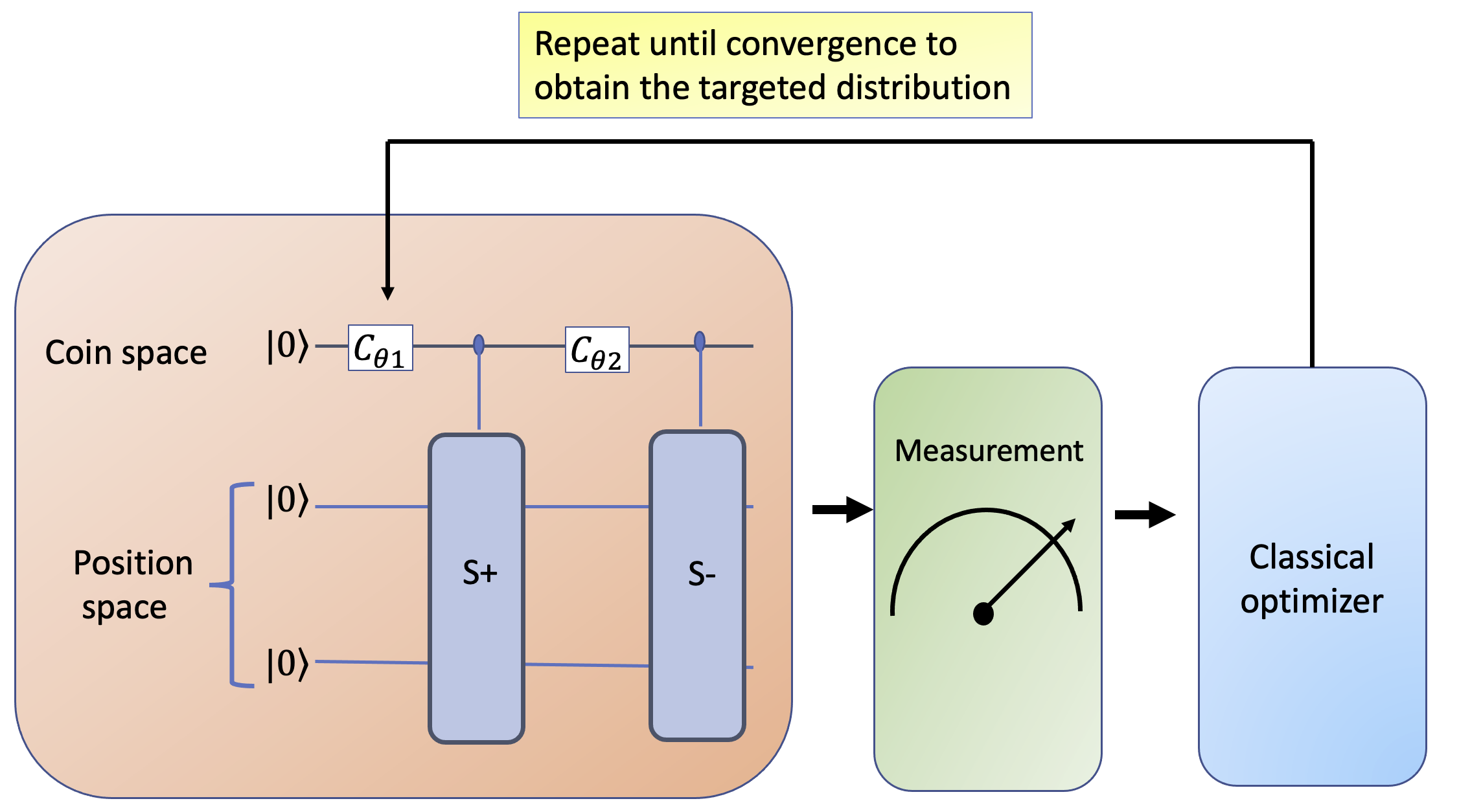}
\caption{ The scheme of loading probability distribution }
\label{fig:FIG4}
\end{figure}

The coin space of an SSQW that performs a controlled motion of a walker on the position space  is similar to the ancilla qubit taking a controlled rotation in Eq.(\ref{eq:controlratation}). The goal is to optimize the coin parameters of an SSQW to achieve the targeted distribution of the position space, and then we only compute the position space. We will accomplish this by using parameterized quantum circuits (PQC) to load the distribution. The steps for this process are as follows:

\begin{itemize}
\item  Begin with a classical targeted data set $p=\{ p_{0},\ldots, p_{2^{N}-1} \} \in R$ sampled from a targeted probability distribution.
\item  Implementation of SSQW uses an ancilla qubit representing the coin space and N qubits representing $2^{N}$ distributions in the position space.
\item Imply $\hat{W}$ operator on the quantum circuit and repeat $t$ steps.
\item Measure the state amplitudes of the position space and compute the trained distribution.
\item Update the coin parameters using the classical optimizer with the mean square error(MSE)
\[ MSE = \frac{1}{2^{N}} \sum^{2^{N}-1}_{i=0} (p^{targeted}_{i}-p^{trained}_{i})^{2} \]
\item Iterate  $k$ times until converge to the targeted distribution
\end{itemize}

\section{Results}

Moreover, a comprehensive simulation to evaluate the performance of the SSQW-based data loading scheme using various distributions was conducted using a standard application programming interface developed by the authors, including normal, log-normal, and actual stock daily return distributions. We demonstrate that this approach has effectively utilized the potential quantum computation in a financial application, as shown by its ability to apply with financial derivative pricing.

\subsection{Performances of various distributions}
 Probability theory\cite{stroock_2010,kallenberg2002foundations}, the mathematical study of randomness, is built upon the concept of probability distributions and the random variables they describe. This article shows examples of two common probability distributions and three actual stock daily return distributions over different periods. The normal distribution, also known as the Gaussian distribution or bell curve, is a typical probability distribution.  It is defined by its mean ($\mu$) and standard deviation ($\sigma$). The normal distribution is often used in statistics and other fields to describe real-world phenomena, such as IQ scores, height, and weight. 

\begin{widetext}
\include{pythonlisting}
\begin{algorithm} [h]
\SetAlgoLined
\KwResult{ Return the trained probability distribution from quantum walk algorithm }
 Initialization: Set the initial parameters for quantum walk and targeted distribution \;
 
 \While{NOT Converged to Targeted Distribution}{
    Uses an ancilla qubit representing the coin space and N-qubits to represent $2^{N}$ distributions in the position space.\;
    Imply $\hat{W}$ operator on the quantum circuit and repeat $t$ steps.\;
    Measure the state amplitudes of the position space and compute the trained distribution.
    Update the coin parameters using the classical optimizer (COBYLA) with the mean square error (MSE)
 }

\caption {Split-Steps Quantum Walk}
\end{algorithm}
\end{widetext}

The log-normal distribution is a probability distribution of a random variable whose logarithm is normally distributed. One of the main properties of the log-normal distribution is that the values are skewed to the right, meaning that the tail of the distribution is on the right side, representing large values. It also has thicker tails than a normal distribution, meaning extreme events are more likely. This is why it helps model variables that fluctuate widely, such as stock prices. The log-normal distribution also can be used to price options by assuming that the underlying asset's price follows a geometric Brownian motion, which means that the logarithm of the asset's price follows a normal distribution. This assumption is the basis of the Black-Scholes model, which is a widely used method for pricing options.

\begin{figure}[htb]
\centering
\subcaptionbox{}{\includegraphics[width=7cm]{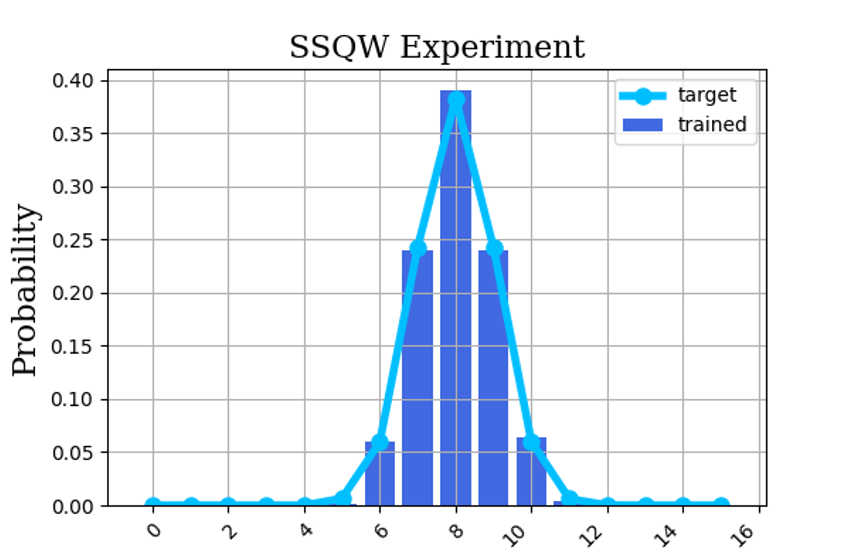}}
\hfill
\subcaptionbox{}{\includegraphics[width=7cm]{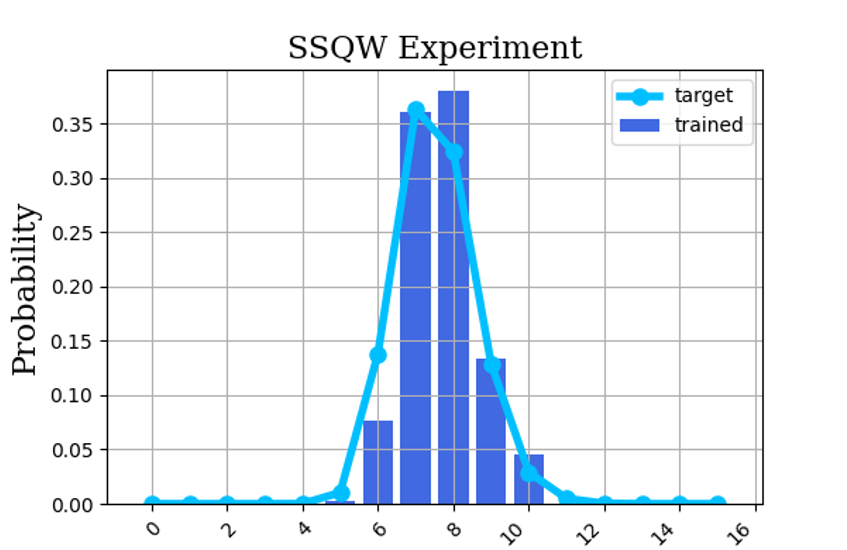}}
\caption{ The performances of (a)Normal distribution and (b)log-normal distribution }
\label{fig:FIG5}
\end{figure}
The methodology begins with generating targeted distributions constructed by drawing $100000$ samples truncated to $(0,15)$ with NumPy. A $5$-qubit quantum circuit is created, with $1$ qubit as the coin space and $4$ qubits as the position space and set the initial state$\Psi_{0}= |0\rangle^{\otimes5}$. The $\hat{W}$ operator has applied $7$ steps, and the state amplitudes of the position space are measured. These measurements are used to compute the trained distribution. The coin parameters are then updated using a classical optimizer and the mean squared error as the objective function. This process is repeated $800$ times to converge to the targeted distribution. The whole process only takes about $20$ seconds. In Fig. \ref{fig:FIG5}, we show better normal and log-normal distribution fitting performances.

Our experience has shown that the initial coin parameters significantly impact the SSQW-based data loading scheme for the normal distribution. However, utilizing the symmetry of the normal distribution allows for simplified implementation and improved results, taking only seconds. This is because the number of coin parameters is reduced from 6 to just 2. On the other hand, implementing the log-normal distribution is difficult as it is skewed to the right and obtaining good results is challenging.

\begin{figure}[htb]
\centering
\subcaptionbox{Apple 2M}{\includegraphics[width=3.7cm]{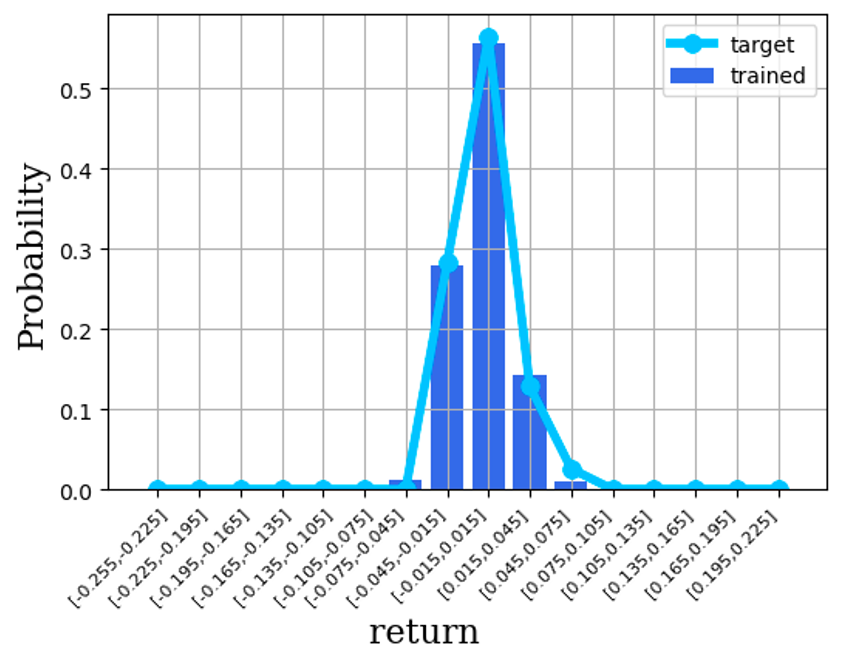}}
\hfill
\subcaptionbox{Apple 3M}{\includegraphics[width=3.7cm]{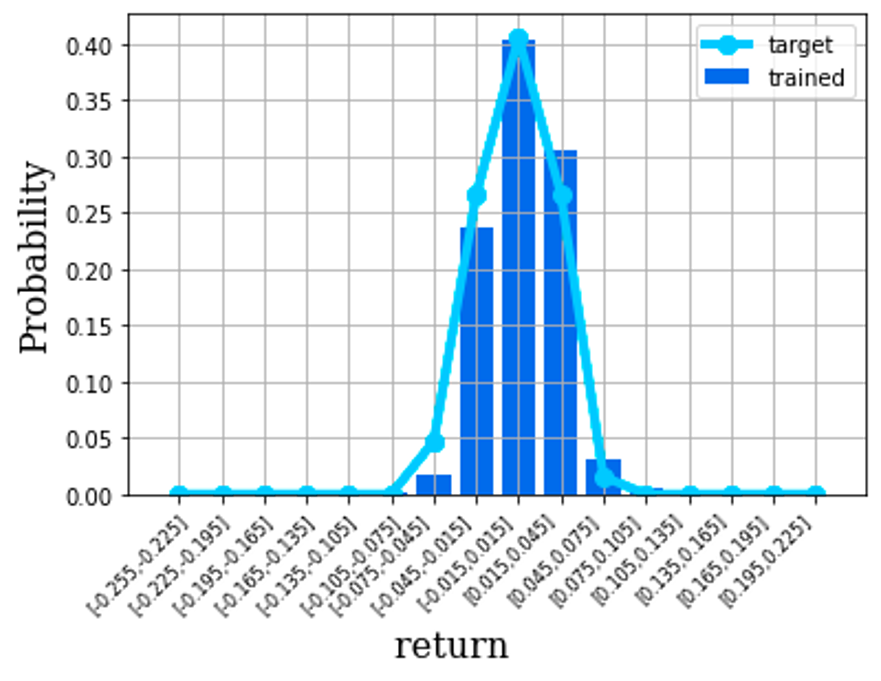}}
\hfill
\subcaptionbox{Microsoft 2M}{\includegraphics[width=3.7cm]{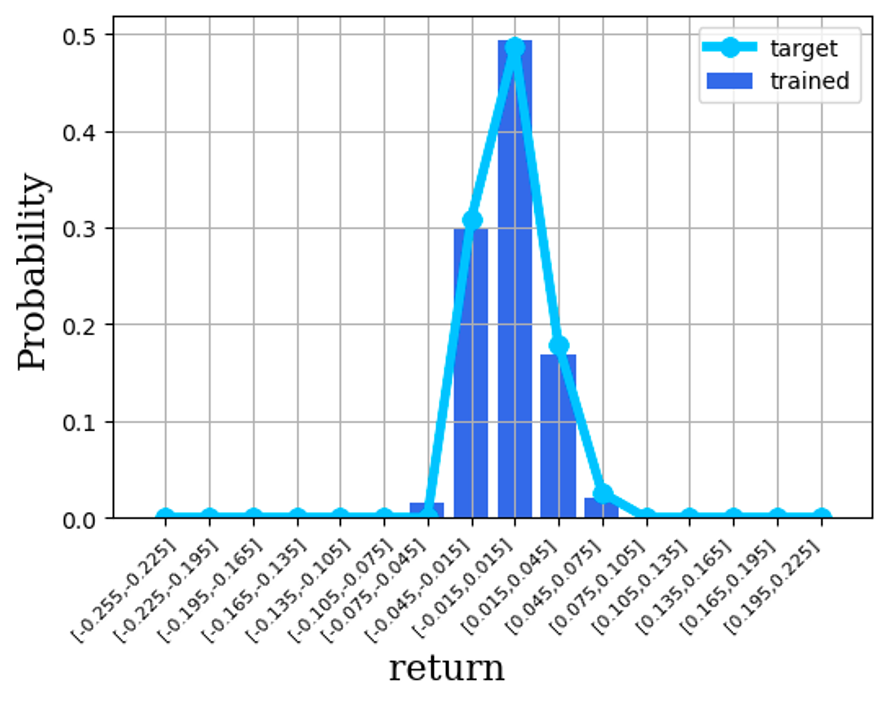}}
\hfill
\subcaptionbox{Microsoft 3M}{\includegraphics[width=3.7cm]{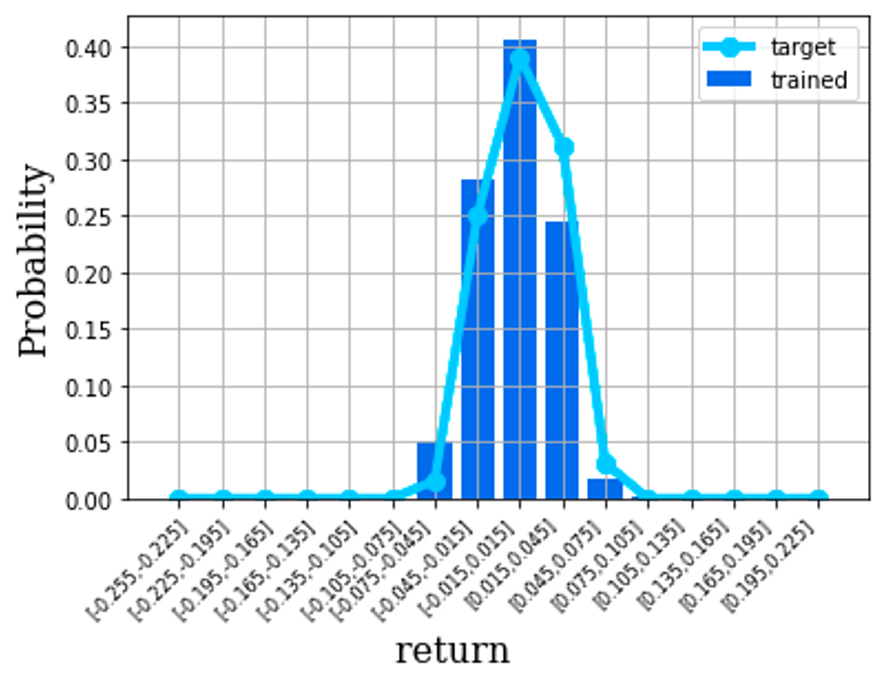}}
\hfill
\subcaptionbox{JP Morgan 2M}{\includegraphics[width=3.7cm]{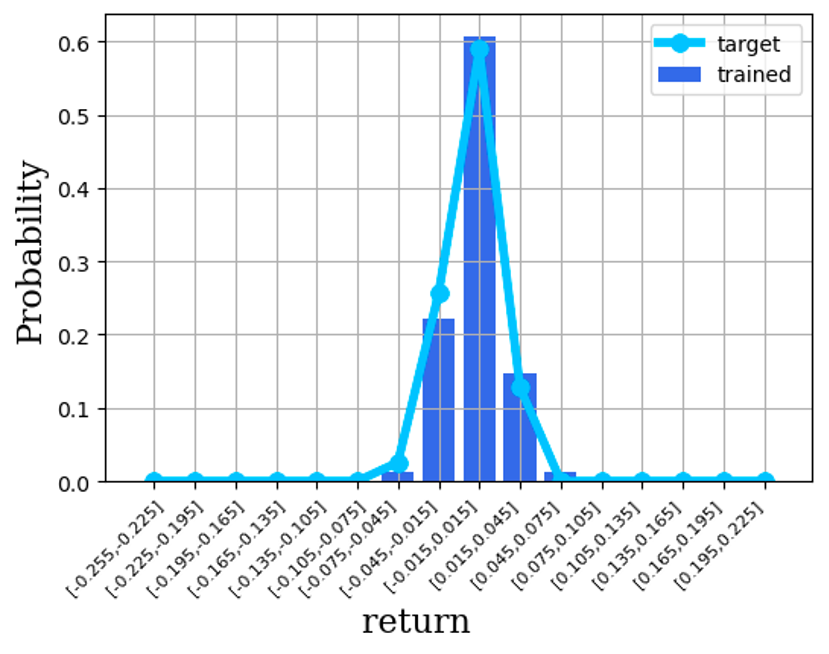}}
\hfill
\subcaptionbox{JP Morgan 3M}{\includegraphics[width=3.7cm]{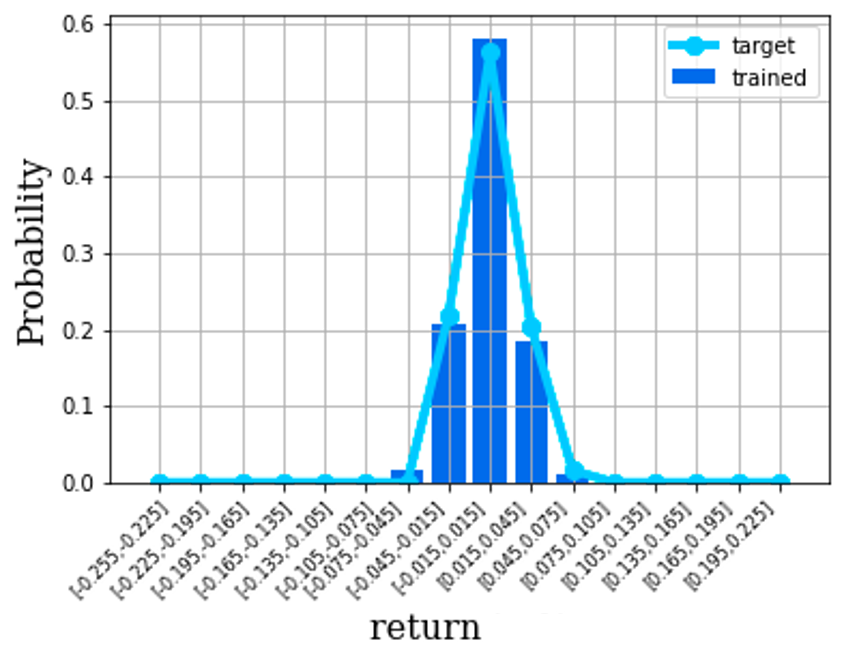}}
\caption{ Exploring Stock Performance Over Different Time Horizons: Daily Returns for Apple, Microsoft, and JP Morgan}
\label{fig:FIG6}
\end{figure}

A daily return distribution is a statistical representation of the daily returns of a financial asset, such as a stock or a commodity. The daily return is the percentage change in the asset's value from one day to the next. The distribution shows how frequently different return values occur and can be used to assess an investment's risk and potential returns. The distribution is typically presented in a histogram, with the x-axis representing the return percentage and the y-axis representing the probability of occurrences. The process of obtaining a daily return distribution starts by collecting real stock data daily from Yahoo Finance over a specified time period and truncates to $(0,15)$
Then we implement the SSQW-based data loading scheme to obtain these results, a more realistic probability distribution of the market shown in Fig \ref{fig:FIG6}. The distributions are similar to the normal distribution with slight skewness. This skewness is important in determining which distribution is appropriate for investment decision-making. 

By implementing the SSQW-based data loading scheme, we are able to obtain results that represent a more realistic probability distribution of the market, as shown in Fig. \ref{fig:FIG6}. The Fig. \ref{fig:FIG6} displays the daily return distributions of APPLE, Microsoft, and JP Morgan over two periods: January 2nd to March 2nd, 2022 (a total of two months), and March 2nd to June 2nd, 2022 (a total of three months). The first column represents the results for the first period and the second column represents the results for the second period. The line chart represents the actual daily stock returns distribution, while the histogram displays the distribution obtained through the SSQW-based data loading methodology.These distributions are similar to the normal distribution with minimal skewness.
The results generated by the SSQW method are highly accurate, with a calculation time of approximately 10 seconds. For financial applications, simulation calculations that are more real-time and offer increased accuracy can be advantageous.

\subsection{Application: European call option price }

Using quantum computing for the pricing problem is still infancy period. More recent works focus on the quantum algorithm for amplitude estimation\cite{Brassard_2002} and Monte Carlo with the pricing of financial derivatives\cite{Rebentrost2018,Zoufal:qGAN2019,Tang2020,Xut2018,Stamatopoulos_2020,Stamatopoulos_2022,Dong_An_2021,Chakrabarti_2021}.
In the following, we demonstrated that the SSQW-based data loading scheme enables the exploitation of the potential quantum advantage in finance, such as European call option pricing.

The Black-Scholes (BS) model\cite{BSmodel}  is widely used in the financial industry to value options and other financial derivatives. is used to determine the theoretical value of an option by using specific parameters such as the underlying asset's price$(S)$, strike price$(K)$, time to expiration$(T)$, risk-free interest rate$(r)$, and volatility$(vol)$. The BS model calculates the option value by assuming that the underlying asset's price follows a geometric Brownian motion, a continuous-time stochastic process. Considering this, the BS model can calculate the probability distribution of the underlying asset's price at expiration. A constant drift$(\alpha)$ and constant volatility $(\sigma)$ characterize the geometric Brownian motion. It has been proven that the natural logarithm of the price of an asset following a geometric Brownian motion will be normally distributed with a mean equal to the drift and a standard deviation similar to the volatility. This is why the BS model is often associated with the log-normal distribution. So the option parameters can be used to calculate the log-normal distribution of the underlying asset's price at expiration. 
The steps to estimate the mean and variance of the log-normal distribution to estimate the volatility $(\sigma)$ and drift $(\alpha)$ of the underlying asset's price, considering non-dividend, initial spot price$(S_{0})$, the daily risk-free interest rate$(r)$, expected return$(\mu)$ and volatility$(vol)$, and the expiration of time $(T)$ are as follows:

\begin{itemize}
\item  Collect historical data on the underlying asset's price, such as daily prices.
\item  Next, Calculate the mean and variance of the logarithm of the asset's prices.
\item Use the mean and variance of the logarithm of the asset's prices to estimate the volatility  $(\sigma)$  and drift $(\alpha)$ of the underlying asset's price. The stock price is log-normally distributed with parameters 
\begin{equation} \label{BS}
\begin{aligned} 
 \sigma &= vol\cdot \sqrt{T}    \\
 \alpha &= \ln{S_{0}} + (\mu-r- \frac{\sigma^{2}}{2} )\cdot T \\ 
\end{aligned}
\end{equation}
\item Finally, using the mean and standard deviation of log-normal distribution we can calculate the underlying asset's price at expiration.The results for estimating $E[max\{ S_{T}-K, 0 \}]$ where $S_{T}$ is the maturity price. 
\end{itemize}

\begin{figure}[htb]
\centering
\subcaptionbox{}{\includegraphics[width=7cm]{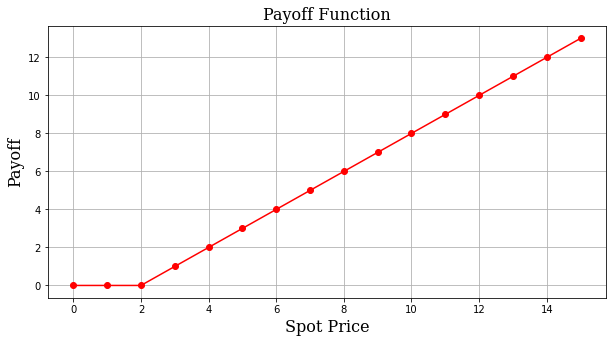}}

\subcaptionbox{}{\includegraphics[width=7cm]{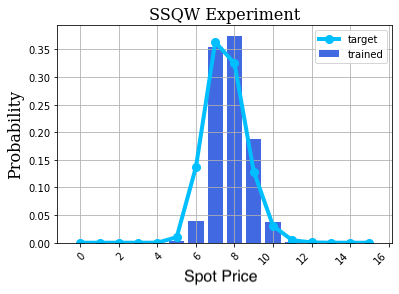}}

\caption{ Demonstration of European call pricing }
\label{fig:FIG7}
\end{figure}

We compare the trained probability distribution with the targeted probability distribution for the parameters $S_{0}=2$,$K=2$, $\sigma = 0.4$, $r = 0.05$ and $T=40$ by plotting them together, as shown in Fig. \ref{fig:FIG7}. Now, the trained probability distribution can be used to evaluate the expectation value of the option's payoff function. We can see that the analytically calculated expected payoff is 5.5342 when using the targeted distribution, and 6.9951 when using the trained distribution. This difference is due to the skewness of the trained probability distribution to the right, leading to a larger expected value than the result obtained from the log-normal distribution of the target. Therefore, when the trained probability distribution is more accurate, resulting in a more favorable outcome after calculation. 

 \section{Discussion}

In this study, we have conducted preliminary theoretical analysis and practical implementation to demonstrate the effectiveness of our approach in addressing probabilistic data loading issues on quantum computers. Our method efficiently loads data while maintaining the accuracy of the probability distribution fitting, with an average computing duration of around 20 seconds, showing great promise for future applications and it allows for the modeling of complex probability distributions. By adjusting the parameters of these coin operations, it is possible to accurately model the probability distribution of the quantum particles' behavior. The SSQW method has the potential to be integrated into larger quantum algorithms, enabling the development of more powerful quantum machine learning and quantum artificial intelligence applications.

Beyond that, there are several areas worth exploring to improve further this method, such as expanding the implementation to more distributions and increasing accuracy. As we have demonstrated, specifying the statistical moments of the probability distribution plays a crucial role in the SSQW-based data loading scheme. For example, the normal distribution can be described with only two parameters, the 1st moment($\mu$) and the 2nd moment($\sigma^{2}$). However, from the point of view of machine learning, the SSQW-based data loading scheme uses six parameters to describe the probability distribution, and it can easily simulate the normal distribution with only two parameters. But when considering probability distributions with more parameters, the SSQW-based data loading scheme may not be able to simulate accurately. Therefore, further research can be done to design the SSQW scheme to widen its applicable restrictions.

Additionally, there is still a substantial improvement that could be made in this study regarding the accuracy of fitting and effectiveness of searching the initial parameter configuration and the future integration with hardware acceleration and machine learning algorithms.  We will refine this method by delving deeper into its details and making it even more sophisticated. 

Furthermore, there is significant potential for enhancing the accuracy of fitting and the effectiveness of searching for initial parameter configurations in this study. The quality of the optimized search parameters is influenced by the initial coin operator, which in turn affects the accuracy of this method. To enhance its efficiency, combining this method with hardware acceleration and machine learning algorithms could yield promising results. Our objective is to refine and further develop this method by thoroughly examining its intricacies and making it even more sophisticated. In conclusion, a method for simulating and predicting future behaviors involves describing a classical probability distribution as a quantum walk. We can then convert this quantum walk into a Hamiltonian\cite{Childs2004,Child_2010} and use it to establish a quantum simulation system. 

Integrating quantum computing and machine learning technologies has immense potential to advance the limitations of machine learning solutions better. An efficient loading scheme for probability distribution, like the SSQW-based data loading scheme, has the potential to be a key component in the development of quantum machine learning and the advancement of quantum artificial intelligence (QAI).  A novel strategy for building SSQW would benefit future researchers to incorporate this solution with future quantum machine learning development. This could allow a deeper understanding of complex phenomena through the quantum walk picture.

\section{ACKNOWLEDGMENTS}
We thank IBM Quantum Hub at NTU for providing computational
resources and accesses for conducting the real quantum device experiments. We acknowledges support from National Science and technology council, Taiwan  under
Grants MOST111-2119-M-033-001, by the research project Applications of quantum computing in optimization and finances.

\bibliographystyle{plain}

\end{document}